\documentclass{article}

\usepackage{PRIMEarxiv}

\usepackage[utf8]{inputenc} 
\usepackage[T1]{fontenc}    
\usepackage{hyperref}       
\usepackage{url}            
\usepackage{booktabs}       
\usepackage{amsfonts}       
\usepackage{nicefrac}       
\usepackage{microtype}      
\usepackage{lipsum}
\usepackage{fancyhdr}       
\usepackage{graphicx}       
\graphicspath{{media/}}     
\usepackage{siunitx}
\usepackage[numbers,sort&compress]{natbib}

\title{Near single-cycle pulse generation using a cascaded all-bulk multi-pass cell compression system}

\author{
  Saga Westerberg, Gaspard Beaufort, Sara Rushe Palacios, Melvin Redon, Chen Guo, Miguel Miranda, \\
  \textbf{Cord L. Arnold, Anne-Lise Viotti} \\
  Department of Physics, Lund University, P.O. Box 118, SE-22100 Lund, Sweden \\
  \texttt{saga.westerberg@fysik.lu.se} \\
}

\begin{document}
\maketitle

\begin{abstract}
We experimentally demonstrate the generation of sub-two-cycle optical pulses using a cascaded post-compression scheme based on two bulk multi-pass cells. Starting from $\SI{220}{\femto\second}$, $\SI{100}{\micro\joule}$ pulses around $\SI{1030}{\nano\meter}$, sequential spectral broadening and pulse compression reduce the pulse duration first to $\SI{50}{\femto\second}$ and ultimately to $\SI{5.5}{\femto\second}$ (1.6 optical cycles) at a central wavelength of $\SI{1058}{\nano\meter}$. The ultra-broadband spectrum is enabled by a dispersion-controlled cavity. Comprehensive spectral, temporal, and spatial characterization confirms excellent pulse quality. Despite operating the second multi-pass cell at peak powers several hundred times above the critical power for self-focusing in fused silica, no significant spatio-spectral or spatio-temporal distortions are observed, enabling direct use of the compressed $\SI{40}{\micro\joule}$ pulses in further experiments. Numerical simulations of the nonlinear spectral broadening show excellent agreement with the experimental results, supporting the underlying physical picture. These findings establish bulk multi-pass cells as an efficient, compact, and robust platform for generating few-cycle pulses with excellent beam quality, offering considerable potential for strong-field and ultrafast applications, including extreme-ultraviolet generation and isolated attosecond pulse production.
\end{abstract}

\section{Introduction}
 
The discovery of high-order harmonic generation\,(HHG) more than 30 years ago led to the development of coherent extreme ultraviolet\,(XUV) light sources with attosecond pulse durations\,\cite{McPherson:87}. While most of these sources deliver trains of attosecond pulses, considerable effort has been devoted to generating trains of only a few attosecond pulses and even isolated attosecond pulses\,(IAP), which enable the tracking of electron dynamics on their natural, attosecond timescale\,\cite{varillas:26}. Beyond temporal resolution, experiments such as coincidence spectroscopy place high demand on statistics and signal-to-noise ratio\,\cite{Osolodkov:20}, making measurements with low repetition rate XUV sources particularly time consuming. Consequently, high repetition rate laser drivers delivering few- to single-cycle pulses\,\cite{Sansone:06}, potentially combined with gating techniques\,\cite{Tcherbakoff:03}, are strongly sought after. 
While ytterbium\,(Yb) lasers can be efficiently operated at kHz to MHz rates, they generate amplified pulses with durations of hundreds of femtoseconds\,(fs) or even picoseconds\,\cite{Troung:25}. To access the few-cycle regime, either optical parametric chirped pulse amplification\,\cite{Fattahi:14} or pulse post-compression schemes can be applied\,\cite{Nagy:21}. In the latter case, the laser pulse undergoes spectral broadening via self-phase modulation\,(SPM) in a Kerr medium, which in turn allows reaching a shorter pulse duration. Examples of such compression platforms are hollow-core fibers\,\cite{Nisoli:24} and multi-pass cells\,(MPCs)\,\cite{Schulte:16}.

Since their introduction for pulse post-compression, MPCs have emerged as a power-efficient and compact approach, consistently delivering high beam quality and excellent spatio-spectral homogeneity\,\cite{Hanna:21,Viotti:22}. They are most commonly realized as off-axis resonators in a Herriott configuration\,\cite{Herriott:64}, allowing for convenient in- and out-coupling of the beam. Spectral broadening via SPM occurs incrementally over many passes in nonlinear Kerr media placed inside the cell, usually noble gases or bulk fused silica, depending on the input peak power. Gas-filled MPCs support input peak powers above several gigawatts, but require operation in vacuum chambers and must be operated below the critical power for self-focusing in the nonlinear medium. Bulk MPCs, on the other hand, can be operated in air, and have been realized with input powers far exceeding the critical power for self-focusing\,\cite{Westerberg:25}. After exiting the cell, the accumulated nonlinear spectral phase needs to be corrected, often using dispersion compensating mirrors or grating compressors. Yb-based sources are now readily compressed from hundreds of femtoseconds, and even picoseconds, to tens of femtoseconds in a single compression stage. 
For example, a $\SI{150}{\femto\second}$ Yb laser system was compressed in a single stage gas MPC to $\SI{15}{\femto\second}$\,\cite{Silletti:25}. 

Both Titanium-sapphire (Ti:Sa) and Yb laser systems have traditionally relied on spectral broadening in hollow capillaries for post-compression to few-cycle pulse durations\,\cite{Nisoli:24}. More recently, however, a sub-$\SI{50}{\femto\second}$ Ti:Sa amplifier was successfully compressed to sub-two-cycle pulse duration using a single gas-filled MPC\,\cite{Daniault:24}. To extend this approach to Yb-based sources, it was proposed that multiple MPC stages could be cascaded to reach the sub-$\SI{10}{\femto\second}$ regime, with the first experimental attempt employing two gas-filled MPCs\,\cite{Balla:20}. To support the bandwidth, the second stage utilized metallic mirrors, to the detriment of beam quality and power efficiency. Subsequently, compression down to $\SI{6.2}{\femto\second}$ was achieved by employing metallic mirrors on water-cooled fused silica substrates in the second stage\,\cite{Muller:21}. Similar results were recently obtained in a three stage scheme for a $\SI{400}{\watt}$ average power laser system\,\cite{Seres:26}. Another alternative is to use ultra-broadband dielectric cavity mirrors with dispersion-controlled coatings\,\cite{Goncharov:23,Rajhans:23,Goncharov:24,Viotti:23}. Among those demonstrations, only one utilized a bulk MPC to reach sub-$\SI{10}{\femto\second}$\,\cite{Viotti:23}. In that work, a $\SI{8.2}{\femto\second}$ pulse was obtained, exhibiting a strong side pulse due to uncompensated third-order dispersion. Spectral broadening towards single-cycle Fourier transform limit was also shown, but spatio-temporal couplings prevented compression. In addition, to the best of our knowledge, the aforementioned few-cycle MPC-based sources employ either silicon or unspecified detectors for spatial evaluation of the ultra-broadband pulses. This makes comprehensive characterization across the full spectral bandwidth centered at the Yb wavelength of $\SI{1030}{\nano\meter}$ challenging as the standard sensitivity window of silicon detectors sharply closes at $\SI{1100}{\nano\meter}$ and the peak quantum efficiency typically occurs between $\SI{700}{\nano\meter}$ and $\SI{900}{\nano\meter}$. 

In this work, we generate post-compressed pulses down to $\SI{5.5}{\femto\second}$ duration, corresponding to 1.6 optical cycles at $\SI{1058}{\nano\meter}$, using a compact two-stage bulk MPC architecture. To our knowledge, this is the shortest pulse duration achieved via an all-bulk MPC system. Each stage utilizes fused silica as the nonlinear medium, the entire compression setup has a footprint of less than $\SI{1}{\square\meter}$ and delivers $\SI{40}{\micro\joule}$ of pulse energy. In the second stage, the MPC employs dispersion compensating mirrors to maintain high peak power throughout propagation in the cell. The output pulses are characterized spatially, temporally and spatio-spectrally, and feature excellent beam and pulse quality. 

The paper is structured as follows: first, we present the experimental setup, then the results from the first compression stage, followed by a study of the spectral broadening in the second stage. Finally, we show the result of the compression of the second stage, including temporal, spatial and spatio-spectral characterization.

\section{Experimental setup} \label{section:2}
\label{expset}
\begin{figure}[h!]
    \centering
    \includegraphics[width=0.95\linewidth]{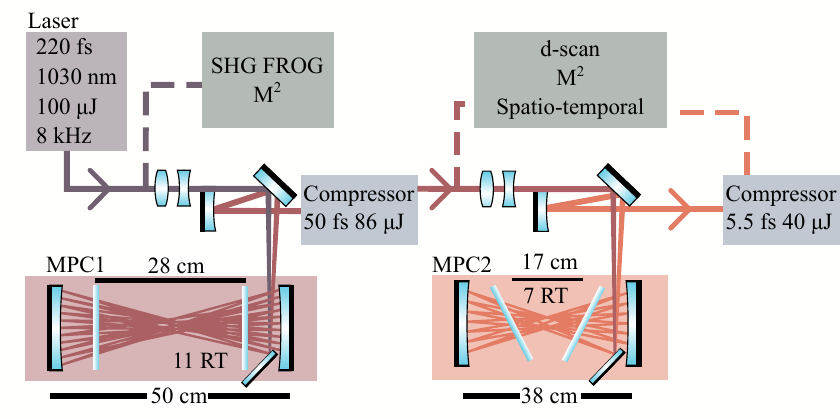}
    \caption{Schematic of the experimental setup showing the laser input parameters, the two cascaded MPC compression stages, available diagnostics and output pulse performances. RT stands for roundtrip and SHG FROG refers to second harmonic generation frequency resolved optical gating.}
    \label{fig:setup}
\end{figure}
The setup is shown schematically in Fig.~\ref{fig:setup}. The laser is an Yb-doped amplified system delivering $\SI{220}{\femto\second}$ full-width at half-maximum\,(FWHM) pulses centered at $\SI{1036}{\nano\meter}$, with an M$^2$ of 1.06\,$\times$\,1.3. The laser is operated at a repetition rate of $\SI{8}{\kilo\hertz}$ (tunable) and up to $\SI{2}{\milli\joule}$ pulse energy. An attenuator consisting of a half-wave plate and a thin film polarizer\,(TFP) is used to control the pulse energy, in this work limited to a maximum of $\SI{100}{\micro\joule}$, before the pulse is sent to the first compression stage. Both compression stages are bulk MPCs\,(MPC1 and MPC2) in a Herriott-type configuration. A mode-matching lens telescope matches the beam to the MPC1 eigenmode, before coupling into the cavity (Fig.~\ref{fig:setup} MPC1). MPC1 consists of two quarter-wave stack dielectric concave mirrors with a radius of curvature\,(ROC) of $\SI{300}{\milli\meter}$ separated by $\sim\SI{500}{\milli\meter}$, and two $\SI{1}{\milli\meter}$-thick anti-reflection coated fused silica plates separated by $\sim\SI{280}{\milli\meter}$, which act as the nonlinear medium where spectral broadening through SPM occurs. After 11 roundtrips the beam exits the cell through the same coupling mirror and is then collimated. Second-order dispersion management is performed with four bounces off two chirped mirrors, compensating for $\SI{-2400}{\femto\second}^2$. 

The compressed output of MPC1 is used as input for MPC2, where the pulse energy is again controlled by an attenuator based on a broadband TFP. Mode-matching for MPC2 is also performed with a lens telescope (see Fig.~\ref{fig:setup} MPC2), and the beam is then coupled into the cell with a metallic mirror. MPC2 consists of a pair of ultra-broadband dispersion-matched dielectric mirrors with ROC $\SI{200}{\milli\meter}$ and separated by $\sim\SI{380}{\milli\meter}$\,\cite{Viotti:23,Rajhans:23}. Two $\SI{1}{\milli\meter}$-thick uncoated fused silica plates separated by $\sim\SI{170}{\milli\meter}$ act as the nonlinear medium. Although ultra-broadband anti-reflection coatings are in principle available, their residual losses remain significant for this application. Therefore uncoated plates are employed and positioned at Brewster angle to minimize reflection losses, resulting in an effective path length of $\sim\SI{1.2}{\milli\meter}$ in each plate. The pair of cavity mirrors is matched to suppress the group delay dispersion\,(GDD) oscillations inherent to broadband chirped mirrors. Each reflection on the MPC mirrors compensates the dispersion introduced by $\SI{3}{\milli\meter}$ of fused silica. This dispersion compensation is necessary to sustain a short pulse in the cavity and maintain the spectral broadening over multiple passes. After 7 roundtrips the beam is coupled out and collimated. Dispersion is managed using a combination of broadband chirped mirrors and materials. We  use a pair of flat mirrors with the same dispersive coating as the MPC2 cavity mirrors. In MPC2, there are actually seven reflections on the mirror opposite of the coupling mirror, and only six on the mirror located behind the coupling mirror, see Fig.~\ref{fig:setup}. Thus, three bounces are introduced via the two matched flat chirped mirrors to compensate for this asymmetry. 
Additionally, two uncoated fused silica wedges are used to fine tune the residual GDD. A $\SI{2.5}{\milli\meter}$ thick potassium dihydrogen phosphate\,(KDP) crystal is used to better balance the third-order dispersion present in the output pulse\,\cite{Silva:14,Miranda:17}. The crystal is cut so that the pulse experiences only the ordinary refractive index upon propagation. Since KDP and fused silica have different ratios of third- to second-order dispersion, combining the two materials provides additional degrees of freedom for compensating the residual dispersion of the MPC2 output pulses. 

The laser output is temporally characterized with a home-built second harmonic generation frequency resolved optical gating (SHG FROG) device\,\cite{Trebino:00}. The compressed output pulses of both MPC1 and MPC2 are measured using a commercial dispersion scan\,(d-scan) device\,\cite{Miranda:26}, designed to handle a wide dispersion range. M$^2$ measurements are performed at each stage. Additionally, spatio-spectral characterization is realized through spatially resolved Fourier transform spectrometry\,\cite{Miranda2014}. 

\section{Results}
\subsection{Characterization of the MPC1 output}
\begin{figure}[h!]
    \centering
    \includegraphics[width=0.85\linewidth]{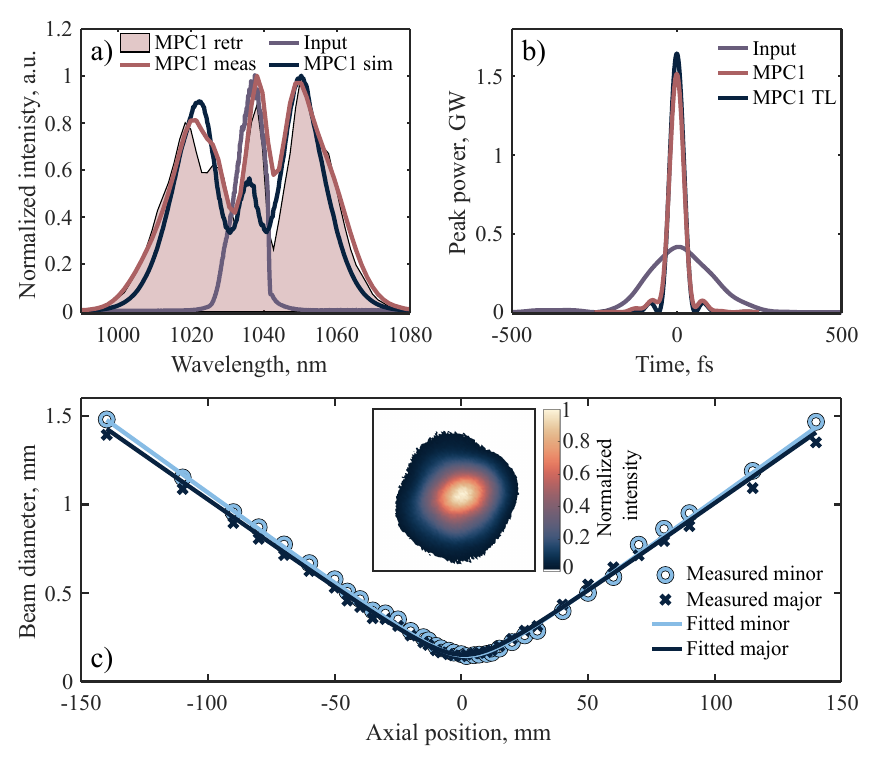}
    \caption{\textbf{MPC1 characterization:} a) Measured input laser spectrum (purple) together with measured (red), simulated (dark blue) and retrieved (shaded area) MPC1 spectra. b) Retrieved pulse profiles of the input laser (purple) and MPC1 (red). The TL pulse profile is shown in dark blue. c) M$^2$ measurement of the MPC1 output. The inset shows the beam profile at focus.}
    \label{fig:Fig2}
\end{figure}

The following results are obtained using an input energy of $\SI{100}{\micro\joule}$ in MPC1. The throughput of the setup including compression is 86\%, resulting in an output pulse energy of $\SI{86}{\micro\joule}$. The losses observed correlate well with the known reflectivity of the cavity mirrors and the fused silica plates.  

Figure~\ref{fig:Fig2} a) shows the measured input spectrum in purple together with the measured output spectrum of MPC1 in red. The simulated spectrum is shown in dark blue and the spectrum retrieved by the d-scan algorithm is displayed as a shaded area. MPC1 is simulated with the SISYFOS\,\cite{Sisyfos} code, using the retrieved input laser pulse as well as the coating reflectivity of the cavity mirrors and fused silica plates. 
The measured, simulated and retrieved spectra show excellent agreement, reproducing similar spectral features. 

Figure~\ref{fig:Fig2} b) displays the retrieved input pulse profile with a duration of $\SI{220}{\femto\second}$ (FWHM), obtained from a SHG FROG measurement. The retrieved pulse profile of the compressed output of MPC1, obtained from a d-scan measurement, is shown in red, and has a FWHM duration of $\SI{50}{\femto\second}$. Additionally, the Fourier transform limited\,(TL) profile is shown in dark blue, with a FWHM of $\SI{49}{\femto\second}$. The input peak power is $\SI{0.4}{\giga\watt}$ and the output peak power is $\SI{1.5}{\giga\watt}$ corresponding to an increase by a factor 3.8. The peak power is obtained by normalizing the retrieved pulse profile by the measured pulse energy, i.e. $\SI{100}{\micro\joule}$ and $\SI{86}{\micro\joule}$ for the input and output pulses, respectively. 
The retrieved pulse profile of the MPC1 output is of high quality, with 90\% of the pulse energy contained within the main pulse. The relative energy content in the main peak is estimated by comparing the integrated power inside a temporal window of twice the FWHM TL pulse duration to the integrated power over a window of $\pm\SI{200}{\femto\second}$. 

In Fig.~\ref{fig:Fig2} c), the result of the MPC1 output M$^2$ measurement is plotted for the minor (light blue circles) and major axes of the beam (dark blue crosses). Minor and major axes refer here to a naming convention describing the corresponding shortest and longest dimensions of the beam in the (x,y) plane. An identical convention is used for the characterization of the input laser beam and MPC2 output beam. For the MPC1 output, the M$^2$ fit yields M$^2$= 1.13 along the major axis and M$^2$= 1.16 along the minor axis, showing preserved spatial beam quality. The inset presents the beam profile at focus.

While we restrict ourselves to operating MPC1 at $\SI{100}{\micro\joule}$ input pulse energy, the setup was tested for up to $\SI{200}{\micro\joule}$ without the need for experimental modifications, resulting in output pulses with duration of $\SI{32}{\femto\second}$ (FWHM) and similar compression quality. The choice to operate with $\SI{50}{\femto\second}$ (FWHM) output pulse duration was made based on simulations of MPC2. Moreover, spatio-spectral measurements were performed at the output of MPC1. No significant spatio-temporal or spatio-spectral couplings were observed, further confirming the pulse post-compression quality. 

\subsection{Spectral broadening in MPC2 and pulse compression to the few-cycle regime}
\begin{figure}[h!]
    \centering
    \includegraphics[width=0.85\linewidth]{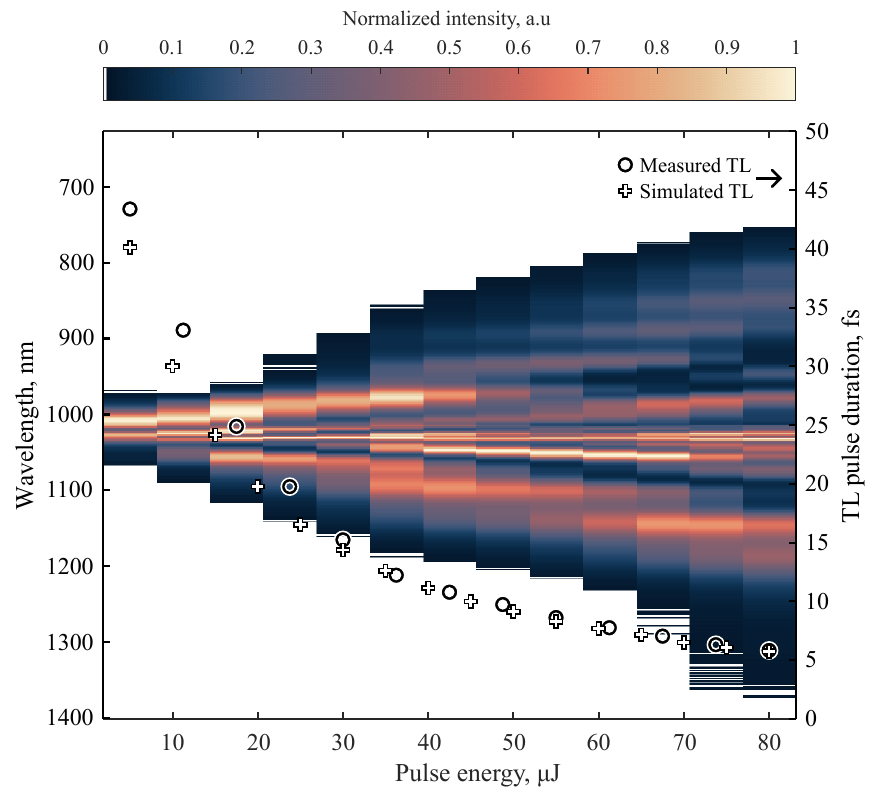}
    \caption{Normalized output spectrum of MPC2, measured for different input pulse energies and corresponding TL pulse durations\,(circles). The crosses represent the TL pulse durations for the simulated spectra.} 
    \label{fig:fig3}
\end{figure}
The attenuator located between MPC1 and MPC2 is used to study the dependence of the spectral broadening on the input pulse energy in MPC2. 
Figure~\ref{fig:fig3} displays the individually normalized spectra measured at the output of MPC2 for different input pulse energies. For each spectrum, the TL pulse duration is plotted\,(circles) as a function of the same energy axis. The crosses depict the TL pulse duration obtained from the SISYFOS simulation of MPC2. The simulation uses the retrieved MPC1 output pulse as input and considers the mirrors' coating reflectivity and dispersion as well as the losses per roundtrip. As the input pulse energy is increased, so is the intensity of the pulse and thereby the spectral broadening due to SPM. The broadest spectrum, obtained at $\SI{80}{\micro\joule}$ input pulse energy, has a TL pulse duration of $\SI{5.3}{\femto\second}$\,(FWHM). The agreement between measured and simulated TL pulse duration is excellent. The slight discrepancy observed at low pulse energies can be explained by small variations in the spectral modulations. 
These differences can be explained by a few assumptions made in the simulations. The input pulse is generated from the retrieved temporal amplitude of MPC1 considering a flat phase and Gaussian beams are used. A realistic scaling factor is introduced to adjust the magnitude of the intensity to mitigate those approximations.  The averaged dispersion of the pair of MPC2 cavity mirrors is employed. Finally, the effect of the different incident angles on the fused silica plates on consecutive passes is not considered as propagation is performed in cylindrical coordinates. Still, the observed spectral broadening is well reproduced by the simulations. 
\begin{figure}[h!]
    \centering
    \includegraphics[width=0.85\linewidth]{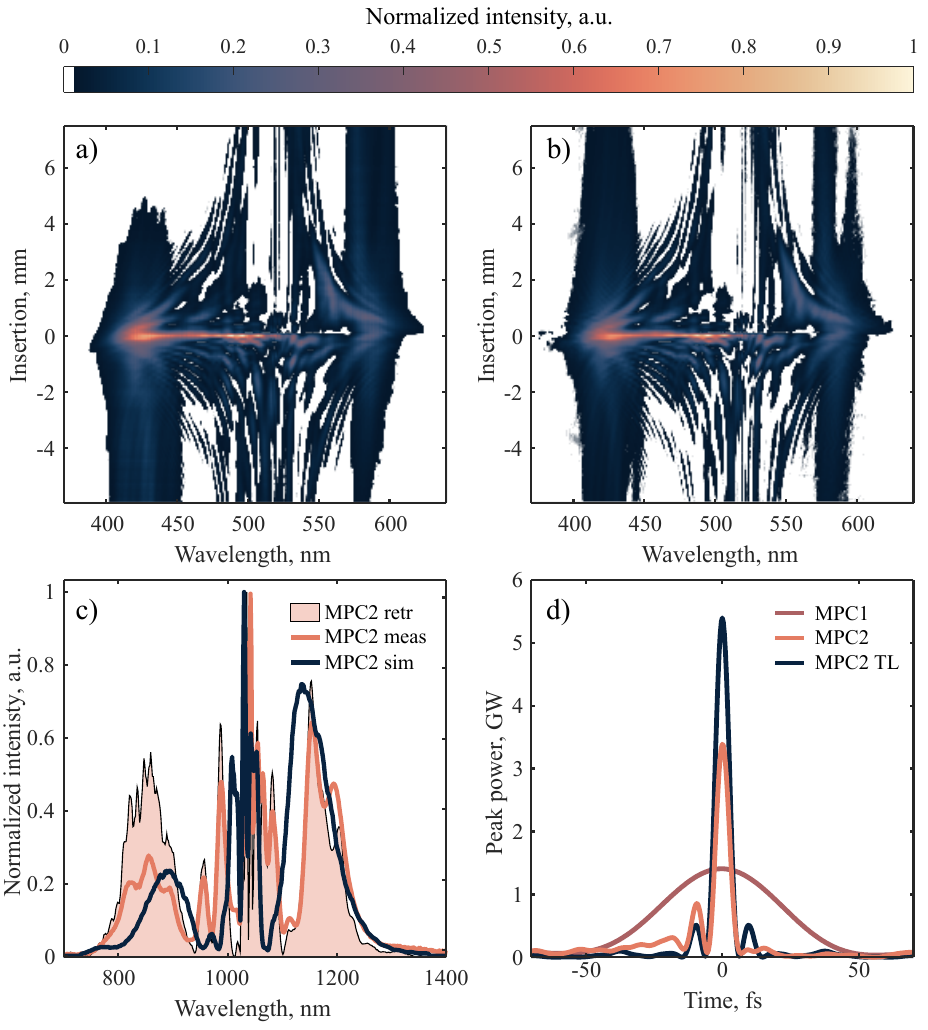} 
    \caption{Measured a) and retrieved b) d-scan traces. c) Measured MPC2 spectrum (orange) together with simulated
(dark blue) and retrieved (shaded area) MPC2 spectra. d) Retrieved pulse profiles of MPC1  (red) and MPC2
(orange). The TL pulse profile for MPC2 is shown in dark blue.  }
    \label{fig:fig4}
\end{figure}

For an input pulse energy of $\SI{80}{\micro\joule}$, the output pulse of MPC2 is compressed using the dispersion management described in section~\ref{section:2}. MPC2 has a transmission of 65\% yielding a pulse energy of $\SI{52}{\micro\joule}$ before dispersion compensation. Most losses are attributed to the uncoated fused silica plates. After dispersion compensation, a pulse energy of $\SI{40}{\micro\joule}$ is measured, corresponding to a compressor transmission of 77\%. The losses in the compressor are mainly due to the uncoated transmission optics, i.e. the fused silica wedges and the KDP crystal.  

Figures~\ref{fig:fig4} a) and b) show the measured and retrieved d-scan traces for the MPC2 output at an input pulse energy of $\SI{80}{\micro\joule}$. In Fig.~\ref{fig:fig4} c) the measured (orange), retrieved (shaded) and simulated (dark blue) spectra are presented. Although the shape of the simulated spectrum varies slightly from the measured one, both spectra exhibit the same TL pulse duration of $\SI{5.3}{\femto\second}$ FWHM. 
In Fig.~\ref{fig:fig4} d), the retrieved pulse profiles from MPC1 (red), MPC2 (orange) and the TL of MPC2 (dark blue) are shown. The retrieved pulse has a FWHM pulse duration of $\SI{5.5}{\femto\second}$, which corresponds to 1.6 optical cycles at a carrier wavelength $\SI{1058}{\nano\meter}$. The MPC2 output pulses have a peak power of $\SI{3.4}{\giga\watt}$. 
To estimate the relative energy content in the main peak of the MPC2 output pulse, we compare the integrated power inside a temporal window of twice the FWHM TL pulse duration to the integrated power over a window of $\SI{200}{\femto\second}$, similar to the case of the MPC1 output. With this integration window, 49\% of the pulse energy is contained in the main pulse, compared to 78\% in the TL case. We observe a low amplitude but extended pedestal in time.

\subsection{Spatial and spatio-spectral characterization of the MPC2 output}
\begin{figure}[h!]
    \centering
    \includegraphics[width=0.85\linewidth]{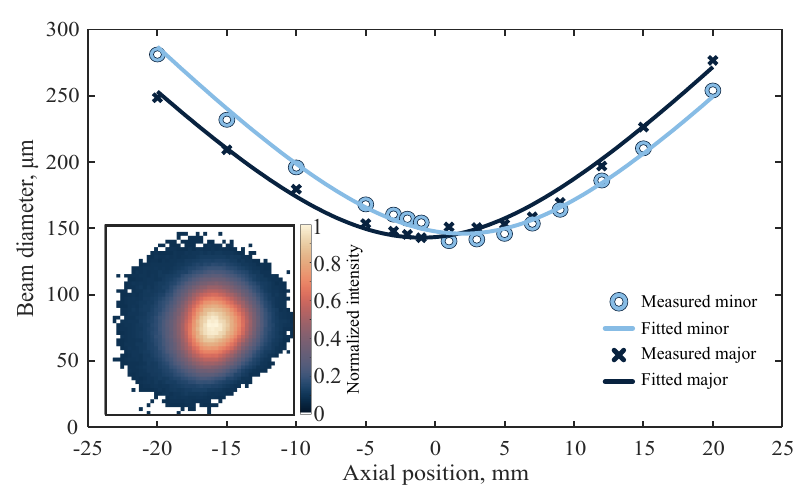}
    \caption{M$^2$ measurement of the MPC2 output beam. The inset shows the measured beam profile at focus.}
    \label{fig:fig5}
\end{figure}
The output spectrum of MPC2 spans from $\SI{750}{\nano\meter}$ to roughly $\SI{1350}{\nano\meter}$. Commonly used silicon-based detectors do not allow to measure spectral content above $\sim\SI{1100}{\nano\meter}$. In order to fully characterize our output beam, we implement M$^2$ and spatio-spectral uniformity measurements using a camera with an indium gallium arsenide\,(InGaAs) sensor\,(\textit{Hamamatsu C16741-40U}).
Figure~\ref{fig:fig5} shows the M$^2$ measurement of the MPC2 output beam and an inset of the beam profile at focus.
The resulting M$^2$ is $1.19\times1.24$, only slightly bigger than for the MPC1 output. 

\begin{figure}[h!]
    \centering
    \includegraphics[width=0.9\linewidth]{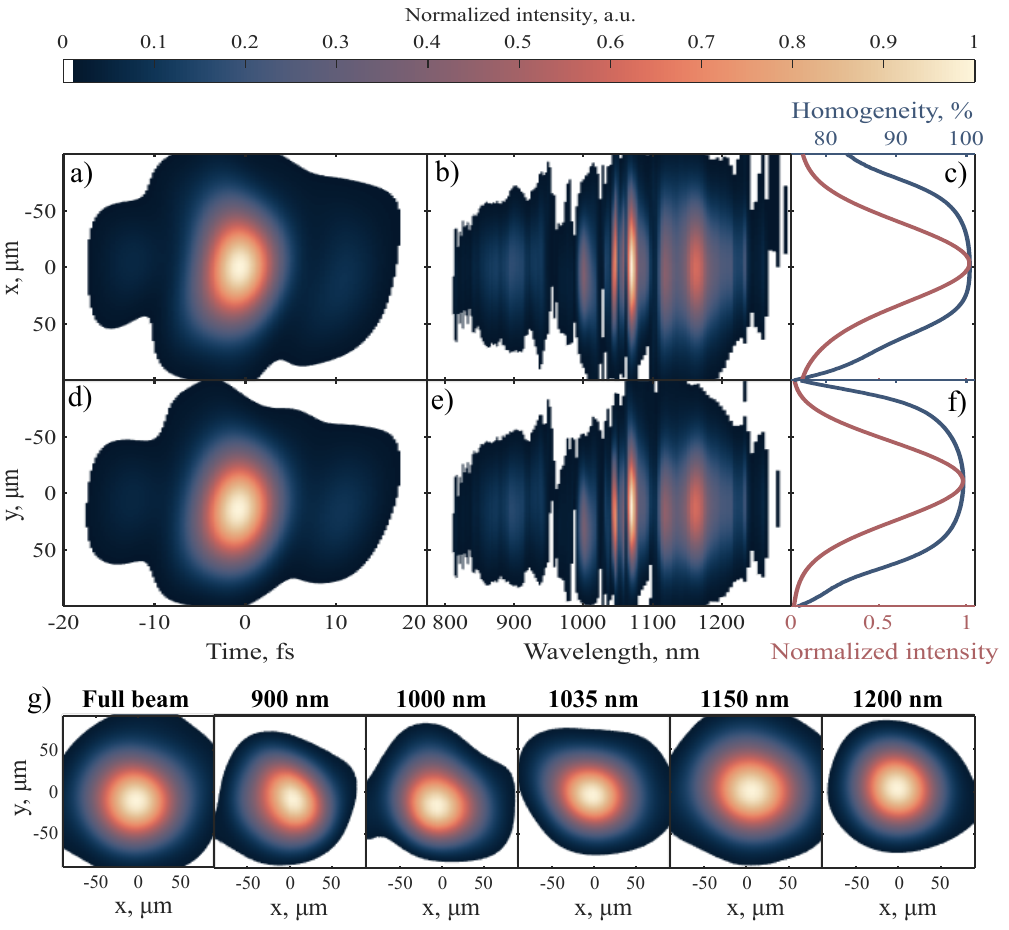}
    \caption{Reconstructed temporal (a) and d)) and spectral (b) and e)) distributions in x and y directions at focus. c) and f) show the spectral homogeneity and beam intensity profile in x and y directions, respectively. g) Retrieved spatial profile at the focus for the full spectrum and at $\SI{900}{\nano\meter}$, $\SI{1000}{\nano\meter}$, $\SI{1035}{\nano\meter}$, $\SI{1150}{\nano\meter}$ and $\SI{1200}{\nano\meter}$. Each beam is individually normalized, and the colorbar applies to all sub-figures. }
    \label{fig:fig6}
\end{figure}
To investigate possible spatio-spectral and spatio-temporal couplings, the output of MPC2 is further characterized using spatially resolved Fourier transform spectrometry\,\cite{Miranda2014}, which is a self-referencing technique. The measurement is performed by splitting the beam into two arms: one arm is unmodified (measurement arm), while the other arm (reference arm) contains a delay stage and a short focal length off-axis parabola, which is used to generate a spherical reference wave. The short focal length of the parabola in combination with sufficient propagation distance, ensures that the reference can be approximated as an ideal spherical wave with a spatially uniform spectrum. Both arms are then recombined and their interferogram is recorded with the InGaAs camera as a function of delay. The spatially resolved spectrum of the output of MPC2 is retrieved through Fourier transform. 
The measurement is performed on the collimated beam, and the retrieved complex field is numerically propagated to the focal plane.

The TL pulse at focus in the x and y directions is shown in Fig.~\ref{fig:fig6} a) and d), respectively. There is no indication of pulse front tilt or other chromatic aberrations in either direction. 
The reconstructed spectra in the x and y directions are displayed in Fig.~\ref{fig:fig6} b) and e), respectively. The spectrum is spatially homogeneous: no significant spatial variation of the spectrum is observed across either direction, indicating absence of angular chirp. From the spatially resolved spectrum, a V-parameter\,\cite{Hanna:21} is calculated to quantify the spatio-spectral homogeneity, shown for the x and y directions in Fig.~\ref{fig:fig6} c) and f), respectively. The average values within the $1/\text{e}^2$ beam width are 98\% in the x-direction and 97\% in the y-direction. 

The retrieved beam profile at focus is shown in Fig.~\ref{fig:fig6} e). Beam profiles are presented at selected wavelengths, individually normalized to facilitate comparison. The wavelength-resolved foci are of similar size and shape, further highlighting the excellent spatio-spectral uniformity of the MPC2 output.

Overall, the spatio-spectral quality of the MPC2 output is high. This implies that MPC2 does not introduce significant spatio-spectral couplings, despite operating MPC2 with input peak powers exceeding the critical power for self-focusing in fused silica by a factor $>300$\,\cite{Boyd:09}. We believe this is the first observation of sub-two-cycle pulses from a bulk MPC driven in the supercritical regime, where, importantly, characterization of the beam and pulse quality over the ultra-broad spectral bandwidth reveals little to no spatio-temporal couplings. This renders the generated few-cycle pulses directly applicable to demanding ultrafast experiments. 


\section{Conclusion}
In this work, we show the post-compression of $\SI{220}{\femto\second}$ pulses from an Yb laser down to $\SI{5.5}{\femto\second}$ FWHM, corresponding to 1.6 optical cycles at a central wavelength of $\SI{1058}{\nano\meter}$. 
The compression factor of 40 is achieved using a compact dual-stage bulk MPC setup, employing dispersion compensating mirrors  in the second stage. 
Temporal, spatial and spatio-spectral characterization is carried out, yielding excellent pulse and beam quality. In particular, an InGaAs camera is employed, allowing us to overcome the upper sensitivity limit of silicon detectors and comprehensively measure our ultra-broadband output beam. Although the second compression stage is operated with a peak power largely exceeding the critical power for self-focusing in the nonlinear medium, the resulting few-cycle pulses are directly suitable for applications. 
The main limitation of the current setup is the remaining losses, mostly due to uncoated optics in the second compression stage, i.e. the nonlinear media and the optical elements used for final dispersion compensation. 
Beyond those practical issues, the pulse energy scalability of the bulk MPC needs to be addressed in order to compete with gas-based MPCs. Multiple strategies can be envisioned, notably spatial shaping\,\cite{Kaumanns:21}, which has been demonstrated recently for bulk MPCs with pulse durations around $\SI{40}{\femto\second}$\,\cite{Koltalo:25}. The question remains to understand if those results carry over to few-cycle pulses and how utra-broadband  spatial shaping can be managed. 

The demonstrated post-compressed few-cycle source constitutes a compact and promising driver for HHG and IAP. Realizing reproducible IAP generation, however, requires carrier-envelope phase (CEP) stability of the MPC-based post-compression system when driven by an industrial-grade Yb-doped laser\,\cite{Rossi:20}. Previous studies have shown that nonlinear compression in MPCs preserves CEP stability\,\cite{Natile:19,Raab:22,Khatri:24}, highlighting the potential of this approach for future attosecond light sources. 

\section*{Funding}

The authors acknowledge the financial support from the Swedish Research Council (Grant Nos. 2021-04691 and 2022-03519) and the Crafoord Foundation. This project was financially supported by the Swedish Foundation for Strategic Research (FFL24-0144). A.-L. V. acknowledges support from Åforsk through the Swedish Foundations’ Starting Grant (RAVIOLI). S.W. and G.B. acknowledge the support of the Helmholtz-Lund International Graduate School (HELIOS, Project No. HIRS-0018).

\section*{Acknowledgments}

We thank G. Arisholm for the theoretical support with simulations and the use of SISYFOS. We thank Anders Persson for fruitful discussion. We thank Hamamatsu Photonics Norden AB for the loan of the InGaAs camera.

\bibliographystyle{ieeetr}  
\bibliography{references}  

\begin{thebibliography}{10}

\bibitem{McPherson:87}
A.~McPherson, G.~Gibson, H.~Jara, U.~Johann, T.~S. Luk, I.~A. McIntyre, K.~Boyer, and C.~K. Rhodes, ``Studies of multiphoton production of vacuum-ultraviolet radiation in the rare gases,'' {\em J. Opt. Soc. Am. B}, vol.~4, pp.~595--601, 1987.

\bibitem{varillas:26}
R.~B. Varillas, P.~Agostini, F.~Ardana-Lamas, C.~L. Arnold, D.~Ayuso, M.~Reduzzi, J.~Benda, J.~Biegert, C.~Bourassin-Bouchet, T.~Brabec, C.~Brahms, A.~C. Brown, D.~Busto, J.~Caillat, F.~Calegari, C.~Callegari, S.~Carlström, Z.~Chang, M.-C. Chen, A.~G. Ciriolo, P.~Corkum, G.~Crippa, R.~de~Q.~Garcia, L.~DiMauro, N.~Dudovich, P.~Eng-Johnsson, D.~Faccialà, P.~Flores, T.~Gadeyne, G.~A. Geloni, C.~Geirger, S.~Gholam-Mirzaei, J.~D. Gorfinkiel, E.~Goulielmakis, M.~Hassan, C.~Hernández-García, P.~Ho, D.~Hui, L.~R. Hutcheson, M.~Ivanov, S.~Kahaly, H.~Kapteyn, N.~Karpowicz, F.~X. Kärtner, M.~Kling, O.~Kneller, D.~H. Ko, P.~M. Kraus, M.~Kubullek, S.~R. Leone, F.~Lépine, A.~L'Huillier, C.-T. Liao, T.~Linker, A.~G. Lohr, M.~Lucchini, L.~B. Madsen, R.~E. Mainz, B.~Major, J.~P. Marangos, D.~Marco, H.~Marroux, S.~Marshallsay, R.~M. Vázquez, R.~Martín-Hernández, Z.~Mašín, M.~Meyer, F.~M. Moreno, M.~Murnane, D.~M. Neumark, M.~Nisoli, M.~Ossiander, S.~Palakka, S.~Patchkovskii, Z.~Pi, L.~Plaja, J.~Poborska, M.~A.
  Porras, K.~C. Prince, D.~N. Purschke, N.~G. Puskar, G.~M. Rossi, J.~R. Rouxel, T.~Ruchon, P.~Rupprecht, P.~Salières, G.~Sansone, F.~Scheiba, M.~Schultze, B.~Schütte, S.~Serkez, M.~A. Silva-Toledo, O.~Smirnova, S.~Stagira, A.~Trabattoni, J.~C. Travers, I.~Tyulnev, M.~Vacher, G.~Vampa, H.~W. van~der Hart, K.~Varjú, A.-L. Viotti, V.~Vishnoi, M.~Vrakking, V.~Wanie, S.~Witte, F.~Xu, V.~S. Yakovlev, L.~Young, D.~Zhu, and C.~Vozzi, ``Roadmap on attosecond science,'' 2026.

\bibitem{Osolodkov:20}
M.~Osolodkov, F.~J. Furch, F.~Schell, P.~Šušnjar, F.~Cavalcante, C.~S. Menoni, C.~P. Schulz, T.~Witting, and M.~J.~J. Vrakking, ``Generation and characterisation of few-pulse attosecond pulse trains at 100 khz repetition rate,'' {\em Journal of Physics B: Atomic, Molecular and Optical Physics}, vol.~53, p.~194003, sep 2020.

\bibitem{Sansone:06}
G.~Sansone, E.~Benedetti, F.~Calegari, C.~Vozzi, L.~Avaldi, R.~Flammini, L.~Poletto, P.~Villoresi, C.~Altucci, R.~Velotta, S.~Stagira, S.~D. Silvestri, and M.~Nisoli, ``Isolated single-cycle attosecond pulses,'' {\em Science}, vol.~314, no.~5798, pp.~443--446, 2006.

\bibitem{Tcherbakoff:03}
O.~Tcherbakoff, E.~M\'evel, D.~Descamps, J.~Plumridge, and E.~Constant, ``Time-gated high-order harmonic generation,'' {\em Phys. Rev. A}, vol.~68, p.~043804, Oct 2003.

\bibitem{Troung:25}
T.-C. Truong, D.~Khatri, C.~Lantigua, C.~Kincaid, and M.~Chini, ``Few-cycle yb-doped laser sources for attosecond science and strong-field physics,'' {\em APL Photonics}, vol.~10, 2025.

\bibitem{Fattahi:14}
H.~Fattahi, H.~G. Barros, M.~Gorjan, T.~Nubbemeyer, B.~Alsaif, C.~Y. Teisset, M.~Schultze, S.~Prinz, M.~Haefner, M.~Ueffing, A.~Alismail, L.~V\'{a}mos, A.~Schwarz, O.~Pronin, J.~Brons, X.~T. Geng, G.~Arisholm, M.~Ciappina, V.~S. Yakovlev, D.-E. Kim, A.~M. Azzeer, N.~Karpowicz, D.~Sutter, Z.~Major, T.~Metzger, and F.~Krausz, ``Third-generation femtosecond technology,'' {\em Optica}, vol.~1, pp.~45--63, Jul 2014.

\bibitem{Nagy:21}
T.~Nagy, P.~Simon, and L.~Veisz, ``High-energy few-cycle pulses: post-compression techniques,'' {\em Advances in Physics: X}, vol.~6, no.~1, p.~1845795, 2021.

\bibitem{Nisoli:24}
M.~Nisoli, ``Hollow fiber compression technique: A historical perspective,'' {\em IEEE Journal of Selected Topics in Quantum Electronics}, vol.~30, no.~6, pp.~1--14, 2024.

\bibitem{Schulte:16}
J.~Schulte, T.~Sartorius, J.~Weitenberg, A.~Vernaleken, and P.~Russbueldt, ``Nonlinear pulse compression in a multi-pass cell,'' {\em Opt. Lett.}, vol.~41, pp.~4511--4514, Oct 2016.

\bibitem{Hanna:21}
M.~Hanna, F.~Guichard, N.~Daher, Q.~Bournet, X.~Délen, and P.~Georges, ``Nonlinear optics in multipass cells,'' {\em Laser \& Photonics Reviews}, vol.~15, no.~12, p.~2100220, 2021.

\bibitem{Viotti:22}
A.-L. Viotti, M.~Seidel, E.~Escoto, S.~Rajhans, W.~P. Leemans, I.~Hartl, and C.~M. Heyl, ``Multi-pass cells for post-compression of ultrashort laser pulses,'' {\em Optica}, vol.~9, pp.~197--216, Feb 2022.

\bibitem{Herriott:64}
D.~Herriott, H.~Kogelnik, and R.~Kompfner, ``Off-axis paths in spherical mirror interferometers,'' {\em Appl. Opt.}, vol.~3, pp.~523--526, Apr 1964.

\bibitem{Westerberg:25}
S.~Westerberg, M.~Redon, A.-K. Raab, G.~Beaufort, M.~Arias~Velasco, C.~Guo, I.~Sytcevich, R.~Weissenbilder, D.~O’Dwyer, P.~Smorenburg, C.~L. Arnold, A.~L’Huillier, and A.-L. Viotti, ``Influence of the laser pulse duration in high-order harmonic generation,'' {\em APL Photonics}, vol.~10, p.~096103, 09 2025.

\bibitem{Silletti:25}
L.~Silletti, A.~b. Wahid, T.~F. Grigorova, L.~Pratolli, A.~Oliveira~e Silva, N.~Hermellin, T.~Mullins, A.~Trabattoni, C.~M. Heyl, C.~Brahms, V.~Wanie, J.~C. Travers, and F.~Calegari, ``Multi-pass cell driven multi-khz dispersive wave emission of broadband deep-uv pulses,'' {\em APL Photonics}, vol.~10, p.~070801, 07 2025.

\bibitem{Daniault:24}
L.~Daniault, J.~Kaur, G.~Gall\'{e}, C.~Sire, F.~Sylla, and R.~Lopez-Martens, ``Sub-2-cycle post-compression of multi-mj energy ti:sapphire laser pulses in a gas-filled multi-pass cell,'' {\em Opt. Lett.}, vol.~49, pp.~6833--6836, Dec 2024.

\bibitem{Balla:20}
P.~Balla, A.~B. Wahid, I.~Sytcevich, C.~Guo, A.-L. Viotti, L.~Silletti, A.~Cartella, S.~Alisauskas, H.~Tavakol, U.~Grosse-Wortmann, A.~Sch\"{o}nberg, M.~Seidel, A.~Trabattoni, B.~Manschwetus, T.~Lang, F.~Calegari, A.~Couairon, A.~L'Huillier, C.~L. Arnold, I.~Hartl, and C.~M. Heyl, ``Postcompression of picosecond pulses into the few-cycle regime,'' {\em Opt. Lett.}, vol.~45, pp.~2572--2575, May 2020.

\bibitem{Muller:21}
M.~Mueller, J.~Buldt, H.~Stark, C.~Grebing, and J.~Limpert, ``Multipass cell for high-power few-cycle compression,'' {\em Opt. Lett.}, vol.~46, pp.~2678--2681, Jun 2021.

\bibitem{Seres:26}
I.~Seres, E.~Shestaev, M.~Tschernajew, P.~Jójárt, C.~Gaida, N.~Walther, T.~Bartyik, B.~Gilicze, Z.~Bengery, D.~Hoff, and et~al., ``Over 400 w average power, sub-two-cycle, carrier-envelope phase-stable fiber laser system,'' {\em High Power Laser Science and Engineering}, vol.~14, p.~e11, 2026.

\bibitem{Goncharov:23}
S.~Goncharov, K.~Fritsch, and O.~Pronin, ``Few-cycle pulse compression and white light generation in cascaded multipass cells,'' {\em Opt. Lett.}, vol.~48, pp.~147--150, Jan 2023.

\bibitem{Rajhans:23}
S.~Rajhans, E.~Escoto, N.~Khodakovskiy, P.~K. Velpula, B.~Farace, U.~Grosse-Wortmann, R.~J. Shalloo, C.~L. Arnold, K.~P. {o}der, J.~Osterhoff, W.~P. Leemans, I.~Hartl, and C.~M. Heyl, ``Post-compression of multi-millijoule picosecond pulses to few-cycles approaching the terawatt regime,'' {\em Opt. Lett.}, vol.~48, pp.~4753--4756, Sep 2023.

\bibitem{Goncharov:24}
S.~Goncharov, K.~Fritsch, and O.~Pronin, ``Amplification-free gw-level, 150 w, 14 mhz, and 8 fs thin-disk laser based on compression in multipass cells,'' {\em Opt. Lett.}, vol.~49, pp.~2717--2720, May 2024.

\bibitem{Viotti:23}
A.-L. Viotti, C.~Li, G.~Arisholm, L.~Winkelmann, I.~Hartl, C.~M. Heyl, and M.~Seidel, ``Few-cycle pulse generation by double-stage hybrid multi-pass multi-plate nonlinear pulse compression,'' {\em Opt. Lett.}, vol.~48, pp.~984--987, Feb 2023.

\bibitem{Silva:14}
F.~Silva, M.~Miranda, B.~Alonso, J.~Rauschenberger, V.~Pervak, and H.~Crespo, ``Simultaneous compression, characterization and phase stabilization of gw-level 1.4 cycle vis-nir femtosecond pulses using a single dispersion-scan setup,'' {\em Opt. Express}, vol.~22, pp.~10181--10191, May 2014.

\bibitem{Miranda:17}
M.~Miranda, J.~{a}o Penedones, C.~Guo, A.~Harth, M.~Louisy, L.~Neori\v{c}i\'{c}, A.~L'Huillier, and C.~L. Arnold, ``Fast iterative retrieval algorithm for ultrashort pulse characterization using dispersion scans,'' {\em J. Opt. Soc. Am. B}, vol.~34, pp.~190--197, Jan 2017.

\bibitem{Trebino:00}
R.~Trebino, {\em Frequency-Resolved Optical Gating: The Measurement of Ultrashort Laser Pulses}.
\newblock Springer, 2000.

\bibitem{Miranda:26}
M.~Miranda, C.~Guo, P.~Guerreiro, C.~Arnold, V.~Amorim, and R.~Romero, ``{dispersion-scan ultrashort pulse characterization implemented with a 4-f pulse stretcher/compressor},'' {\em Optica Open}, 1 2026.

\bibitem{Miranda2014}
M.~Miranda, M.~Kotur, P.~Rudawski, C.~Guo, A.~Harth, A.~L'Huillier, and C.~L. Arnold, ``Spatiotemporal characterization of ultrashort laser pulses using spatially resolved fourier transform spectrometry,'' {\em Opt. Lett.}, vol.~39, no.~17, pp.~5142--5145, 2014.

\bibitem{Sisyfos}
G.~Arisholm and H.~Fonnum, ``Simulation system for optical science (sisyfos) - tutorial, version 2,'' {\em Tech. Rep., Ser. FFI-rapport. Norwegian Defense Research Establishment (FFI)}, 2021.

\bibitem{Boyd:09}
R.~W. Boyd, S.~G. Lukishova, and Y.~R. Shen, {\em Self-focusing: past and present}.
\newblock Springer, 2009.

\bibitem{Kaumanns:21}
M.~Kaumanns, D.~Kormin, T.~Nubbemeyer, V.~Pervak, and S.~Karsch, ``Spectral broadening of 112 mj, 1.3 ps pulses at 5 khz in a lg10 multipass cell with compressibility to 37 fs,'' {\em Opt. Lett.}, vol.~46, pp.~929--932, Mar 2021.

\bibitem{Koltalo:25}
V.~Koltalo, S.~Westerberg, M.~Redon, G.~Beaufort, A.-K. Raab, C.~Guo, C.~L. Arnold, and A.-L. Viotti, ``Energy scaling in a compact bulk multi-pass cell enabled by laguerre–gaussian single-vortex beams,'' {\em APL Photonics}, vol.~10, p.~040801, 04 2025.

\bibitem{Rossi:20}
G.~M. Rossi, R.~Mainz, Y.~Yang, F.~Scheiba, M.~Silva-Toledo, S.~Chia, P.~Keathley, S.~Fang, O.~M{\"u}cke, C.~Manzoni, G.~Cerullo, G.~Cirmi, and F.~K{\"a}rtner, ``Sub-cycle millijoule-level parametric waveform synthesizer for attosecond science,'' {\em Nature Photonics}, vol.~14, pp.~629--635, Oct. 2020.

\bibitem{Natile:19}
M.~Natile, A.~Golinelli, L.~Lavenu, F.~Guichard, M.~Hanna, Y.~Zaouter, R.~Chiche, X.~Chen, J.~F. Hergott, W.~Boutu, H.~Merdji, and P.~Georges, ``Cep-stable high-energy ytterbium-doped fiber amplifier,'' {\em Opt. Lett.}, vol.~44, pp.~3909--3912, Aug 2019.

\bibitem{Raab:22}
A.-K. Raab, M.~Seidel, C.~Guo, I.~Sytcevich, G.~Arisholm, A.~L'Huillier, C.~L. Arnold, and A.-L. Viotti, ``Multi-gigawatt peak power post-compression in a bulk multi-pass cell at a high repetition rate,'' {\em Opt. Lett.}, vol.~47, pp.~5084--5087, Oct 2022.

\bibitem{Khatri:24}
D.~Khatri, T.-C. Truong, C.~Lantigua, C.~Kincaid, M.~Britton, and M.~Chini, ``Carrier-envelope phase-stabilized ultrashort pulses from a gas-filled multi-pass cell,'' {\em Appl. Phys. Lett.}, no.~9, 2024.

\end{thebibliography}

\end{document}